\documentclass[3p,numbers, twocolumn]{elsarticle}
\usepackage{amsmath,amssymb, underscore}
\pdfoutput=1
\usepackage{subfig,float}
\usepackage{multicol}
\usepackage{graphicx}
\usepackage{color}
\usepackage{xspace}
\biboptions{numbers,sort&compress}
\shortcites{CollaborationEtAlPlanck2015}
\usepackage{hyperref}
\hypersetup{  pdfauthor={...},  pdftitle={...},  pdfsubject={...},  urlcolor=blue,}

\newcommand{\ee}{e}
\newcommand{\Mp}{M_{ \mathrm{Pl}}}
\newcommand{\rhoRe}{\rho_{\rm re}}
\newcommand{\Tre}{T_{\rm re}}
\newcommand{\Nre}{N_{\rm re}}
\newcommand{\wre}{w_{\rm re}}

\begin{document}

\renewcommand{\figurename}{Fig.}

\title{Reheating constraints on K\"ahler Moduli Inflation}

\author[du]{R. Kabir\corref{cor1}}
\ead{rakesh.kabir@nic.in }

\author[du]{A. Mukherjee\corref{cor2}}
\ead{am@physics.du.ac.in }

\author[du]{D. Lohiya\corref{cor3}}
\ead{dlohiya@physics.du.ac.in }

\cortext[cor1]{Corresponding author}

\address[du]{Department of Physics and Astrophysics, University of Delhi, Delhi-110007, India}


\begin{abstract}

The end of inflation is connected to the standard cosmological scenario through reheating. During reheating, the inflaton oscillates around the minimum of the potential and thus decays into the daughter particles that  populate the Universe at later times. Using cosmological evolution for observable CMB scales from
the time of Hubble crossing to the present time, we translate the constraint on the spectral index  $n_s$ from Planck data to the constraint on the reheating scenario in the context of K\"ahler Moduli Inflation. We find that the details of the potential is irrelevant if the analysis is done strictly within the slow-roll formalism. In addition, we extend the de-facto analysis generally done only for the pivot scale to all the observable scales which crossed the Hubble radius during inflation. We study how the maximum number of e-folds varies for different scales, and the effect of the equation of state and potential parameters.
\end{abstract}

\maketitle

\section{ Introduction}\label{sec:intro}
Inflation is regarded as the standard cosmological paradigm to describe the physics of the very early Universe which tackles several problems in cosmology, particularly the question of the origin of cosmological structures. It leads to a causal mechanism to generate almost scale invariant fluctuations on cosmological scales, with small deviations that follow from the precise microphysics of inflation. This prediction is consistent with the measurements
of the cosmic microwave background (CMB) anisotropies by the Planck satellite. The latest data allow us to constrain the inflationary model besides giving a slightly red tilted spectral index \(n_s\) = 0.9682 $\pm $ 0.0062, ruling out exact scale invariance \(n_s\) = 1 at over 5$\sigma $~\citep{Ade2013,Collaboration2013}. 

In the evolution of the Universe, reheating is a transition epoch connecting the end of inflation to radiation and subsequently matter dominated stages of the Universe. Reheating is important as it is the responsible mechanism for the origin of matter in the Universe~\cite{KofmanOrigin1996}. During reheating the vacuum energy of the inflaton field gets transferred to the decay particles. Reheating models are classified broadly in two classes, perturbative and non-perturbative. In simple models, Perturbative decay involves a three-point interaction term of the form $\sigma\phi\chi^2$ where $\phi$, $\chi$ and $\sigma$ are the inflaton, some (scalar) decay product and coupling constant with dimensions of mass, respectively \citep{AminEtAlNonperturbative2015}. Non-perturbative models include parametric resonance decay~\cite{KofmanEtAlReheating1994,KofmanEtAlTowards1997}, tachyonic instability~\citep{GreeneEtAlInflaton1997,ShuhmaherEtAlNon2006,DufauxEtAlPreheating2006,AbolhasaniEtAlTachyonic2010,FelderEtAlTachyonic2001}, and instant preheating\cite{FelderEtAlInstant1999}. 

Just after the inflation ends, the Universe is in the non-thermal preheating state~\cite{AminEtAlNonperturbative2015}. Over time the distribution of states are covered by scattering events. This creates a blackbody spectrum characterized by a final temperature $\Tre$, at which the radiation-dominated era begins.   

To study reheating in an inflationary scenario for some given interactions between the inflaton and other matter fields, an effective equation of state (eos) has to be derived numerically. In general, the eos is a function of time, $\wre(t)$, and changes as the Universe evolves during the various stages of reheating. During this evolution $\wre(t)$ takes values from $-1/3$ at the end of inflation to $0$ in the initial phase to $1/3$  at the start of the radiation dominance. The densities at the end of inflation and at the end of reheating are related by
\begin{equation}
\ln\frac{\rhoRe}{\rho_{end}}=-3\int_{a_{end}}^{a_{re}}(1+\wre)\frac{da}{a}.
\end{equation}\label{eq:teos}
Since the analytic integration of the right hand side may not be possible for the arbitrary time dependent $\wre$, it is often considered to be constant throughout the entire reheating era. However in the recent literature, time-varying eos has also been considered~\cite{FreitasEtAlCMB2015} (also see~\cite{LozanovEtAlEquation2016} for a related analysis). 

As the fluctuations generated during reheating remain sub-horizon, they cannot leave an observable imprint at the level of the CMB or LSS~ \cite{CookEtAlReheating2015}. Therefore it is difficult to constrain the reheating era, except in some speculative scenarios~\cite{TaruyaEtAlCosmological1998,BassettEtAlMassless2000,FinelliEtAlParametric2000,TsujikawaEtAlWhen2002,ChambersEtAlLattice2008,BondEtAlNon2009,BassettEtAlGeneral1999,BassettEtAlMetric1999,BassettEtAlRestoring2000,BethkeEtAlAnisotropies2013,EastherEtAlSupersymmetry2014}. Still we can put lower and upper bounds on the reheating temperature from the considerations of primordial nucleosynthesis and the energy scale of inflation~\cite{SteigmanPrimordial2007}.  

This work is organized as follows: in Sec.~\ref{sec:formalism} we detail the derivation of the reheating duration and of the temperature at the end of reheating as a function of the spectral index, for canonical single-field inflationary models and for reheating scenarios that can be described in terms of a constant effective equation of state; in Sec.~\ref{sec:poly} we review the analysis of\cite{DaiEtAlReheating2014}, see also~\cite{CookEtAlReheating2015}, for a power law potential and we discuss the constraints from reheating on the inflationary parameters; in Secs.~\ref{sec:kahler} we compute the relations between inflationary and reheating parameters in the K\"ahler inflation models and discuss the bounds placed on some of these models by reheating; in Sec.~\ref{sec:conclusion} we present our conclusions.

\section{ Basic Formalism}\label{sec:formalism}
To set up our notation, we briefly review the formalism in \cite{CookEtAlReheating2015} and~\cite{UenoEtAlConstraints2016} to constrain the the $e$-folding number of reheating 
$N_{\rm re}$ and the reheating temperature $T_{\rm re}$ using an observational 
constraint on the spectral index $n_s$.

For a single-field inflation model with a potential $V(\phi)$, 
the inflaton obeys
\begin{eqnarray}
&&\ddot\phi+3{\dot a\over a}\dot\phi+{\partial V\over \partial \phi}=0,
\label{eq:fielddot}
\end{eqnarray}
where the dot indicates differentiation with respect to
cosmic time, and $a$ is the scale factor
determined by the Friedman equation:
\begin{eqnarray}
&&\left({\dot a\over a}\right)^2
  ={1\over 3M_{\rm pl}^2}\left({\dot\phi^2\over 2}
  +V(\phi)\right).
\label{eq:Friedman}
\end{eqnarray}
Adopting the slow-roll approximation during inflation,
the above equations are approximated as
\begin{eqnarray}
&&3H\dot\phi+V'(\phi)=0,\label{eq:appfieldddot}
\\
&&H^2={V(\phi)\over 3M_{\rm pl}^2},\label{eq:appFriedman}
\end{eqnarray}
where the prime denotes differentiation with respect to $\phi$, and 
$H=\dot a/a$ is the Hubble parameter. Using the definitions of the potential slow-roll parameters 

\begin{equation}
\epsilon  \equiv   \frac{1}{2}M_{\text{pl}}^2 \left(\frac{V_{\phi }}{V}\right)^2\text
\, , 
\end{equation}

\begin{equation}
\eta    \equiv M_{\text{pl}}^2 \left(\frac{V_{\phi \phi }}{V}\right) 
\end{equation}

 the scalar spectral index is expressed as
\begin{equation}\label{eq:ns}
n_s= 1 - 6 \epsilon  + 2 \eta
\end{equation}
and the tensor-to-scalar ratio as
\begin{equation}
r_T= 16 \epsilon \, .
\end{equation}

Using the definition of energy density $\left({\dot\phi^2/ \left(2 {M_{\rm pl}^2}\right)} +V(\phi)\right)$ and the slow-roll approximated Eqs.~\eqref{eq:appfieldddot} and \eqref{eq:appFriedman}, the energy density during the inflation epoch can be  written as $\rho=(1+\epsilon/3)V$.
We take the end of inflation as $\epsilon=1$, at which the energy density of the Universe can be written as 
\begin{eqnarray}
  \rho_{\rm end} ={4\over 3}V(\phi_{\rm end})={4\over 3}V_{\rm end},
\label{eq:rhoend}
\end{eqnarray}
where $\phi_{\rm end}$ is the value of the scalar field at the end of inflation.

Finally the number of e-folds between horizon crossing of a perturbation of wavenumber $k$ and the end of inflation  is given by

\begin{equation}\label{eq:Nk}
N_k \equiv \ln\biggl({a_{\rm end}\over a_{k}}\biggr) 
=-\frac{1}{M_{\text{pl}}^2}\int_{\phi _{end}}^{\phi _k} (\frac{V}{V_{\phi }}) \, d\phi
\end{equation}
where \(\phi _{end, k }\) are the values of $\phi $ at the end of inflation and when mode $k$ leaves the Hubble radius.

Let us assume that during the reheating epoch, the Universe is dominated by an energy component due to the scalar field with an effective 
equation-of-state parameter $w_{\rm re}$ which is taken to be constant during the entire reheating period. At the end of the reheating epoch, the reheating temperature is related to the energy 
density of the Universe as
\begin{eqnarray}
  \rho_{\rm re}={\pi^2 g_{\rm re}\over 30} T_{\rm re}^4,
\label{eq:rhore}
\end{eqnarray}
where $T_{\rm re}$ is the reheating temperature, and $g_{\rm re}$ is the
number of relativistic species at the
end of reheating, which can be assumed to be $g_{\rm re}={\cal O}(100)$.
Defining the scale factor at the end of reheating, $a_{\rm re}$, we can write the $e$-folding number of the reheating epoch,
\begin{multline}\label{eq:Nredef}
N_{\rm re} \equiv
\ln\biggl({a_{\rm re}\over a_{\rm end}}\biggr)
=\\-{1\over 3(1+w_{\rm re})}
\ln\biggl(\frac{\rho_{\rm re}}{\rho_{\rm end}}\biggr),
\end{multline}
where $a_{\rm re}$ is the scale factor at the end of reheating.

Now we show how an observational constraint on the spectral index $n_s$ can constrain the $e$-folding number $N_{\rm re}$.
This depends on the model, i.e. potential $V(\phi)$, which sometimes (depending upon the expression for potential) makes the analytically writing $\Nre$ in terms of $n_s$ very difficult. The derivation of $\Nre$ hinges on the relation $a_kH_k=k$, which is satisfied by a perturbation mode with the wavenumber $k$, where $a_k$ and $H_k$ are respectively the scale factor and Hubble parameter just before the mode $k$ exits Hubble radius during the inflation. In cases where $\Nre$ and $n_s$ both can be written as function of $\phi_k$, a parametric plot can be obtained between $\Nre$ and $n_s$, otherwise numerical techniques have to applied. Then the logarithmic version of the relation $a_kH_k=k$ can be broken into different epochs
\begin{eqnarray}
0=\ln\biggl({k\over a_kH_k}\biggr)=\ln\biggl({a_{\rm end}\over a_k}{a_{\rm re}\over a_{\rm end}}
{a_0\over a_{\rm re}}{k \over a_0 H_k}\biggr),
\label{NNN0}
\end{eqnarray}
where $a_0$ is the scale factor at the present epoch. Using the definitions Eqs.~(\ref{eq:Nredef}) and (\ref{eq:Nk}), Eq.~(\ref{NNN0}) yields
\begin{eqnarray}
N_k+N_{\rm re}+\ln\biggl({a_{0}\over a_{\rm re}}\biggr)+\ln\biggl({k\over a_0H_k}\biggr)=0. 
\label{NNN2}
\end{eqnarray}

Using the conservation of entropy, we may write
\begin{eqnarray}
{a_{\rm re}\over a_0}=\left(\frac{43}{11g_{re}}\right)^{{1}/{3}}{T_0\over T_{\rm re}},
\label{eq:entropyCons}
\end{eqnarray}
where $T_0=2.725$~K is the temperature of the Universe at the present epoch. Using 
Eq.~(\ref{eq:rhore}), $\Tre$ can be replaced with $\rhoRe$  in Eq.~(\ref{eq:entropyCons})
\begin{eqnarray}
{a_{\rm re}\over a_0}=\left(\frac{43}{11g_{re}}\right)^{{1}/{3}}{T_0}
\left({\pi^2 g_{re}\over 30 \rho_{re}}\right)^{1/4}. 
\label{2.5.X}
\end{eqnarray}

However $\rhoRe$ can be written in terms of $V_{end}$ and $\Nre$ using connection between them using Eq.~\eqref{eq:rhoend}. 
Therefore, using Eqs.~(\ref{eq:rhoend}) and (\ref{eq:Nredef}), we have
\begin{multline}
  \rho_{re}={4\over 3}V_{\rm end}\left({a_{\rm re}\over a_{\rm end}}\right)^{-3(1+w_{\rm re})}\\=
{4\over 3}V_{\rm end}e^{-N_{\rm re}3(1+w_{\rm re})}.
\end{multline}
Then, the logarithm of Eq.~(\ref{2.5.X}) yields the following expression in terms of $N_{\rm re}$: 
\begin{multline}
\ln\left({a_{\rm re}\over a_0}\right)={1\over 3}\ln\left(\frac{43}{11g_{re}}\right)
+{1\over 4}\ln\left({\pi^2 g_{re}\over 30}\right)\\
+{1\over 4}\ln\left({3T_0^4\over 4 V_{\rm end}}\right)+{3N_{\rm re}(1+w_{\rm re})\over 4}. 
\label{2.5.XX}
\end{multline}

Using the amplitude of the scalar perturbations, $A_s=H^4/(4\pi^2\dot\phi^2)$,
and the slow-roll approximation, we may write
\begin{eqnarray}
H_{k}=\frac{\pi M_{\rm pl}\sqrt{r A_s}}{\sqrt{2}}.
\label{eq:Hkdef}
\end{eqnarray}
Inserting Eqs.~(\ref{2.5.XX}) into Eq.~(\ref{NNN2}), we finally have
\begin{multline}
N_{\rm re}=\frac{4}{1-3w_{\rm re}}\Bigg[-\ln\left(\frac{k}{a_{0}T_{0}}\right)\\-\frac{1}{4}\ln\left(\frac
{30}{\pi^2g_{\rm re}}\right)-\frac{1}{3}\ln\left(\frac{11g_{re}}{43}\right)
-\ln\left(\frac{V_{\rm end}^{1/4}}{H_k} -N_{k}\right)\Bigg]. 
\label{eq:Nre}
\end{multline}

In our analysis, we use $n_{s}=0.9682\pm0.0062$
and $k=0.05~{\rm Mpc}^{-1}$ as  pivot wavenumber, and $T_0 =3.522 \times 10^{25} {\rm Mpc}^{-1}$ as present temperature of the Universe in the cosmological units. we adopt 
the amplitude of the scalar perturbation at the pivot scale $A_s$ 
given by 
$A_{s}=2.196 \times 10^{-9}$ 
(Table 4 of \cite{CollaborationEtAlPlanck2015}). Small variations in $A_{s}$ have negligible effects on reheating predictions. 
We use $n_{s}=0.9682\pm0.0062$. 
The last two terms in Eq.~\eqref{eq:Nre} are model dependent. As $g_{\rm re}$ can be safely taken to be of order of 100, for a particular wave number, a simple expression for $\Nre$ can be obtained for Planck's pivot scale $k=0.05~{\rm Mpc}^{-1}$:
\begin{multline}
N_{\rm re}=\frac{4}{1-3w_{\rm re}}\Bigg[61.6
-\ln\left(\frac{V_{\rm end}^{1/4}}{H_k}\right)-N_{k}\Bigg]. 
\label{eq:Nrepivot}
\end{multline}
As we will plot $\Tre$ on the base-$10$ logarithmic scale, qualitatively it will be negative scaled version of the $\Nre$ plot.    
Combining Eqs.~(\ref{eq:rhoend}), (\ref{eq:Nre}), and (\ref{eq:entropyCons}), we also have~\cite{EshaghiEtAlCMB2016}
\begin{eqnarray}
T_{\rm re}=\mathrm{exp}\Bigg[-\frac{3}{4}(1+w_{\rm re})N_{\rm re}\Bigg]
\left(\frac{2V_{\rm end}}{5\pi^2}\right)^{1/4}.
\label{2.5.8}
\end{eqnarray}
Because the wavenumber $k$ and $n_s$ are related implicitly through the scalar field $\phi$, using $H_ka_k=k$, Eqs.~(\ref{eq:ns}) and (\ref{eq:Nk}), 
one can write $N_{\rm re}$ and $T_{\rm re}$ as
functions of the spectral index $n_s$. 

In previous works~\cite{CookEtAlReheating2015,UenoEtAlConstraints2016} considering constraints on reheating, the analysis has been done only for the pivot scale $k=0.05~{\rm Mpc}^{-1}$. Considering $\Nre$ as a function of wavenumber $k$, we have extended this to all other observable scales too, particularly to find out what is the maximum possible $\Nre$ for each observable scale. To this end, from Eq.~\eqref{eq:Nre}, $\Nre$ becomes a function of $k$ in addition to $n_s$. In particular, $V_{end}$ and $N_k$ and $H_k$ will be treated as functions of $n_s$ through inverting the function $\phi_k(n_s)$ whereas dependence on $k$ will arise through the term $\ln\left(\frac{k}{a_{0}T_{0}}\right)$. For pivot scale $k=0.05~{\rm Mpc}^{-1}$, the above logarithmic term plus two other model independent terms in Eq.~\eqref{eq:Nre} makes 61.6~\cite{CookEtAlReheating2015}. So, in essence, we have generalized previously used $\Nre(n_s)_{k=0.05~{\rm Mpc}^{-1}}$ to $\Nre(n_s,k)$, where to get maximum $\Nre$, $n_s$ takes the lower bound or the upper bound value from the constraint on the $n_s$, i.e.
\begin{align}
\begin{cases} 
\textmd{Max }\Nre(n_s = 0.9620, k )  \quad\quad \text{for} \quad \wre < 1/3\,,\\
\textmd{Max }\Nre(n_s = 0.9744, k )  \quad\quad \text{for} \quad \wre > 1/3\,.
\end{cases}
\end{align}
From the analytic expression of $\Nre$ in Eq.~\eqref{eq:Nre}, it can be inferred that maximum possible $\Nre$ will depend logarithmically on wavenumber $k$, but to get the variation in it with the order of magnitude, we will have to perform numerical calculations for the specific potentials. 

\section{Polynomial Potential}\label{sec:poly}

Before moving to the K\"ahler potential, let us consider a polynomial type potential
\begin{align}\label{eqqq1}
V = \frac{1}{2} m^{4 - \alpha} \phi^{\alpha}.
\end{align}
This was considered in the context of reheating in \cite{MartinEtAlEncyclopaedia2013,DaiEtAlReheating2014,MartinEtAlObserving2015}. This model is analytically tractable.\\ 
The number of e-folds between the time the pivot scale exited the Hubble radius and the end of inflation can be derived using Eq.~(\ref{eq:Nk})
\begin{align}
N_k = \frac{1}{2 \alpha M_P^2} \left(\phi_k^2 - \phi_{end}^2 \right) .
\end{align}\label{eq:Nkpoly}
The potential in these polynomial models is generally steep enough so that $\phi_k \gg \phi_{end}$, and therefore it is appropriate to approximate
\begin{align}\label{eq:NKpolyapp}
N_k \approx \frac{1}{2 \alpha M_P^2} \phi_k^2\, .
\end{align}
Notice that the above approximation may not be valid for a general potential. 
To write $N_k$ as a function of $n_s$, we observe from Eq.~(\ref{eq:ns}) that  $n_s$ is a function of $\phi$. Therefore by inverting $n_s(\phi_k)$ into $\phi_k(n_s)$ and using that in  Eq.~(\ref{eq:NKpolyapp})  one finds 
\begin{align}
N_k = \frac{\alpha + 2}{2 (1 - n_{s})}.
\end{align}
From Eq.~(\ref{eq:Hkdef}) and using the previous equation, $H_k$ is given by
\footnote{ There were typographical errors in Eqs.(3.5) and (3.6) of Ref.~\cite{CookEtAlReheating2015}, which have been corrected here.}
\begin{align}
H_k =\pi M_{P}\sqrt{\frac{4\alpha A_{s}}{\alpha+2}(1-n_{s})}.
\end{align}
If $V_{\rm end}(\phi)$ is used as such then we have to know the value of the inflaton mass $m$. 
However owing to the $V \propto \phi^{\alpha}$ (model dependent part) and $V \propto H^2$ (from slow-roll Friedmann equation~\eqref{eq:appFriedman}), we can write 
\begin{equation}
V_{\rm end}(\phi) = 3 M_{\rm pl}^2 H_{k}^2 \frac{\phi_{end}^{\alpha}}{\phi_k}
\end{equation}
Lastly one computes $V_{end}$ in terms of $n_s$ and $A_s$,%
\begin{multline}\label{eqqq2}
V_{end}= 6 \pi^2 M_P^4 A_s \frac{1}{2^{\alpha /2}} (1- n_s) \left (\frac{\alpha (1- n_s)}{ (\alpha + 2)} \right)^{(\alpha +2)/2},
\end{multline}
where the value of the inflaton field at the end of inflation $\phi_{end}$ follows from the condition $\epsilon = 1$.\\
Thus $N_k$, $H_k$, and $V_{end}$ are all expressed as functions only of $\alpha$, $n_s$ and $A_s$ and one may plot $N_{re}$ (and $T_{re}$) as a function of $n_{s}$ for some fixed values of $w_{re}$ and $\alpha$. 

\begin{figure}[H] 
\begin{center}
\subfloat[]{\includegraphics[height=60mm,width=70mm]{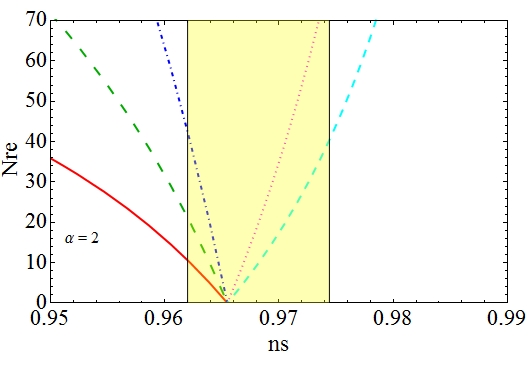} \label{F:Nrepoly2}}\\
\subfloat[]{\includegraphics[height=60mm,width=70mm]{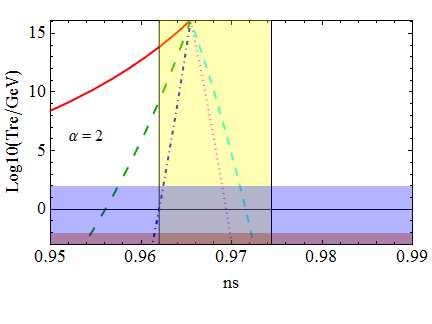} \label{F:Trepoly2}}
\caption{ (color online). Plots of $N_{re}$ and $T_{re}$, the length of reheating and the temperature at the end of reheating respectively, for polynomial potentials with exponent $\alpha$. The solid red line corresponds to $w_{re} = -1/3$, the dashed green line to $w_{re} = 0$, the dot-dashed blue line to $w_{re} = 1/6$, the dotted magenta line to $w_{re} = 2/3$ and the dot-dashed black line to $w_{re} =1$. The yellow shaded region corresponds to the $1 \sigma$ bounds on $n_s$ from Planck.  In \protect\subref{F:Trepoly2} temperatures below the light purple shaded region are ruled out by BBN. The light purple pink region is below the electroweak scale, assumed 100 GeV for reference. This region is not disallowed but would be interesting in the context of baryogenesis.}
\label{F:poly2}
\end{center}
\end{figure}

\begin{figure}[H]
\begin{center}
\subfloat[]{\includegraphics[height=60mm,width=70mm]{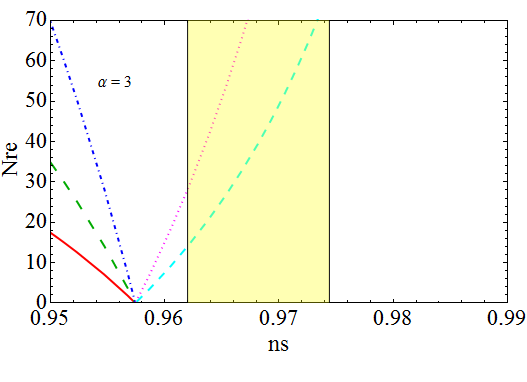} \label{F:Nrepoly3}}\\
\subfloat[]{\includegraphics[height=60mm,width=70mm]{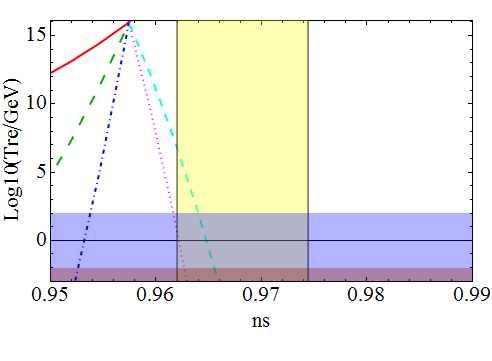} \label{F:Trepoly3}}
\caption{ (color online). Same as for figure \ref{F:poly2} but here we have taken  $\alpha = 3$.\label{F:poly3}}
\end{center}
\end{figure}

We take instantaneous reheating as the limit $N_{re} \rightarrow 0$, visualized in the $N_{re}$ and $\Tre$ plots as the point where all the $w_{re}$ lines meet. Such instantaneous reheating leads to the maximum temperature at the end of reheating, and the equation of state parameter $w_{re}$ is irrelevant.
We plot in Fig.~\ref{F:poly2} $N_{re}$ and $T_{re}$ predictions for $\alpha = 2$  and 3. As the instantaneous reheating is lying near the central value of $n_s$, $\alpha =2$ give promising predictions when compared with the Planck data. As explained in Ref.~\cite{CookEtAlReheating2015}, for $\alpha = 2/3$, our polynomial potential takes the form which is favored by axion-monodromy models.  As the instantaneous reheating point comes out of the yellow strip in Fig.~\ref{F:poly3}, the cases $\alpha =3$ and $\alpha =4$ (not plotted for the reason of space) are difficult to reconcile with $w_{re}\leq 1$ even if one considers the 2$\sigma$ bounds on $n_{s}$\footnote{{Warm inflation is an exception where $\phi^{4}$ may still be viable~\cite{BartrumEtAlimportance2014,Bastero-GilEtAlObservational2014}.}}. \\


\section{K\"ahler Potential}\label{sec:kahler}
These models are string inspired models and appear in type IIB string
theories from Calabi-Yau compactification. KMII scenarios
have been derived and studied in \cite{ConlonEtAlKahler2006,BondEtAlRoulette2007,KrippendorfEtAlMetastable2009,Blanco-PilladoEtAlKahler2010,KawasakiEtAlKahler2011,LeeEtAlKahler2011}. If internal spaces
are weighted projective spaces, one of the K\"ahler moduli can be considered 
as an inflaton field and its potential, in the large field
limit, reads
\begin{equation}\label{eq:kmii:pot}
V(\phi)=M^4\left(1-\alpha\frac{\phi}{\Mp}\ee^{-\phi/\Mp}\right) ,
\end{equation}
where $\alpha$ is a positive dimensionless parameter.

\begin{figure}[H]
\begin{center}
{\includegraphics[width=75mm]{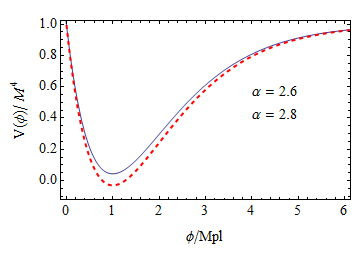}}
\caption{ An illustration of the potential for different values of  $\alpha$.}\label{Fig:potKM2}
\end{center}
\end{figure}

This potential depends on one free parameter,
$\alpha$. As explained in \cite{MartinEtAlEncyclopaedia2013}, in order for slow-roll inflation to occur, the range of possible values for
$\alpha$ is restricted. Within the narrow allowed range, the predictions of the model is almost independent of $\alpha$. The potential~(\ref{eq:kmii:pot})
for two values of $\alpha$ is displayed in Fig.~\ref{Fig:potKM2}. The potential blows up at $\phi=0$, and then decreases as $\phi$ increases. It reaches a minimum at $\phi=\Mp$. Note that the minimum is lower for $\alpha=2.8$ than for $\alpha=2.6$. The asymptotic value $V=M^4$ is reached when $\phi\rightarrow
+\infty$. As originating from the string theory, the potential is derived under the large field assumption. Therefore inflation is assumed to proceed from the right to the left, 
in the increasing $\phi>\Mp$ branch of the potential. It is to be observed that the
minimum value of the potential is located at $\phi=\Mp$ and is
$V_{\mathrm{min}}=M^4\left(1-\alpha/e\right)$. Therefore, for the potential to be positive definite everywhere, the condition $0<\alpha<e\simeq 2.72$ must be satisfied. However, this condition is relevant only when one considers both the increasing ($\phi>\Mp$) as well as decreasing ($\phi<\Mp$) branch of the potential, and can be safely ignored if one restricts the field excursion to the increasing branch only. 

\subsection{Slow-roll parameters}
\begin{figure}[H]
\begin{center}
\subfloat[]{\includegraphics[height=60mm,width=70mm]{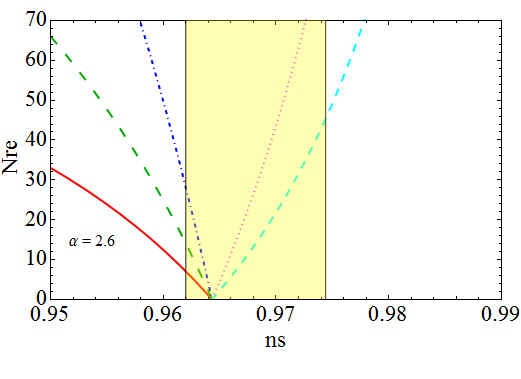} \label{F:kmiiNre26}}\\
\subfloat[]{\includegraphics[height=60mm,width=70mm]{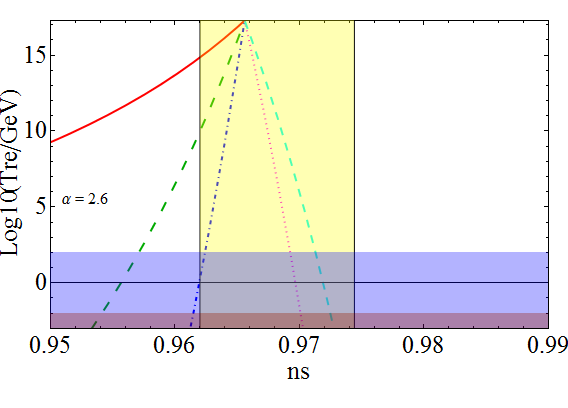} \label{F:kmiiTre26}}
\caption{ (color online). Plots of $\Nre$ and $\Tre$, the length of reheating and the temperature at the end of reheating respectively, for K\"ahler potential with exponent $\alpha = 2.6$.\label{F:kmii26}}
\end{center}
\end{figure}

\begin{figure}[H]
\begin{center}
\subfloat[]{\includegraphics[height=60mm,width=70mm]{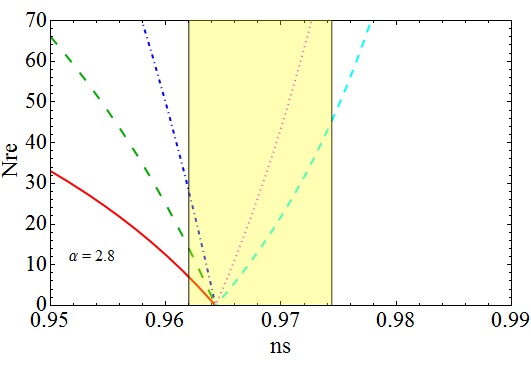} \label{F:kmiiNre28}}\\
\subfloat[]{\includegraphics[height=60mm,width=70mm]{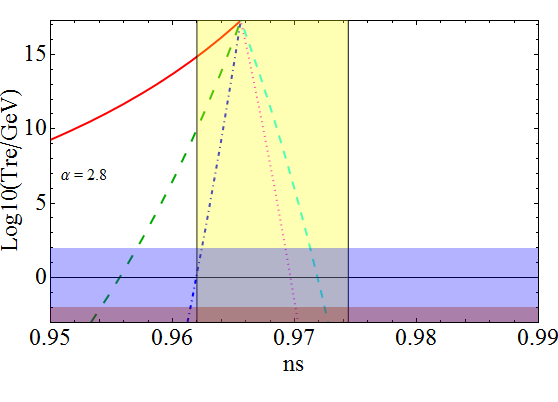} \label{F:kmiiTre28}}
\caption{ (color online). Same as for figure \ref{F:kmii26} but here we have taken  $\alpha = 2.8$. \label{F:kmii28}}
\end{center}
\end{figure}

The slow-roll parameters for the K\"aher potential can be expressed as
\begin{equation}
\epsilon(\phi) = \frac{\alpha ^2 (\Mp-\phi )^2}{2 \left(\Mp \ee^{\frac{\phi }{\Mp}}-\alpha  \phi \right)^2},
\end{equation}\label{eq:epskm2}

\begin{equation}
\eta(\phi)=\frac{\alpha  (\phi -2 \Mp)}{\alpha  \phi -\Mp \ee^{\frac{\phi }{\Mp}}},
\end{equation}\label{eq:etakm2}
and therefore using \eqref{eq:ns}, we get 
\begin{equation}
n_s=\frac{\Mp \left(\alpha ^2 (2 \phi -3 \Mp)+4 \alpha  \ee^{\frac{\phi }{\Mp}} (\Mp-\phi )+\Mp \ee^{\frac{2 \phi }{\Mp}}\right)}{\left(\Mp \ee^{\frac{\phi }{\Mp}}-\alpha  \phi
   \right)^2}.
\end{equation}\label{eq:nskm2}
As $n_s$ depends on $\phi$ in a complex way and can not be inverted to get $\phi(n_s)$, we will have to do numerical simulation, whose results are shown in Figs.~\ref{F:kmii26} to \ref{F:kmii28}. 

For $\wre$ = 1/3,assuming $g_{re} = 100$, and Planck's pivot scale, Eq.~\eqref{eq:Nre} implies
\begin{align}\label{Eq:nrewre13}
 61.6 =   \ln \left(\frac{ V_{end}^{\frac{1}{4}}}{H_{k}} \right) + N_{k} \,.
\end{align}

Using Planck's central value for $A_s$, Eq.~\eqref{Eq:nrewre13} gives specific predictions for $n_s$
\begin{equation}
\begin{cases}
n_s = 0.964225   \quad\quad \text{for} \quad\alpha = 2.6\,,\\
n_s = 0.964250   \quad\quad \text{for} \quad\alpha = 2.7\,,\\
n_s = 0.965277   \quad\quad \text{for} \quad\alpha = 2.8\,.\\
\end{cases}
\end{equation}

\begin{figure}[H]
\begin{center}
\subfloat[]{\includegraphics[height=60mm,width=70mm]{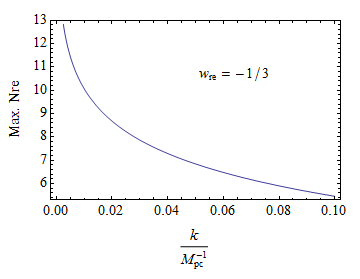} \label{F:kmiiNremaxvska28r1}}\\
\subfloat[]{\includegraphics[height=60mm,width=70mm]{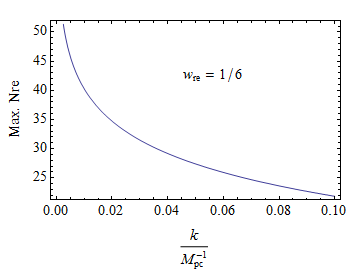} \label{F:kmiiNremaxvska28r3}}
\caption{ (color online). For $\alpha=2.8$, the variation in maximum $\Nre$ versus $k$ when \protect\subref{F:kmiiNremaxvska28r1} $\wre =-1/3$ and  $\wre = 1/6$, respectively.\label{F:kmiiNremaxvska28}}
\end{center}
\end{figure}

Fig~\ref{F:kmiiNremaxvska28} shows the variation of maximum $\Nre$ versus wave number $k$ for fixed $\wre$ and $\alpha$. As already pointed in Sec.~\ref{sec:formalism}, the variation is logarithmic in nature and the effect of $\wre$ is that of scaling the y-axis. For a fixed $\wre$, there was minor variation in the plots for $\alpha=2.8$ and $\alpha=2.6$, obviously due to minor horizontal variation in the convergent point for the potential(see Fig.~\ref{F:kmii26} and~\ref{F:kmii28}).  

\section{ Discussion of Results}

As is apparent from the plots of $N_{re}$ and $\Tre$ in \cite{CookEtAlReheating2015,UenoEtAlConstraints2016} and the present work for different potentials, qualitatively the $w_{re}$ curves are same for different potentials due to slow-roll approximation. In other words we can say that slow-roll approximation overrides the differences in potential shapes for $N_{re}$ (or $\Tre$) vs $n_s$ curves. These characteristic curves are due to the solution being on the slow-roll trajectory. Further, we make few general observations on the $N_{re}$ and $\Tre$ plots. The role of single parameter $\alpha$ is just to shift the group of $w_{re}$ curves  horizontally, whereas each $w_{re}$ curves are tilted to the right or the left depending upon whether $w_{re}$ is less  or greater than $\frac{1}{3}$, respectively. Apparently, a $w_{re} = \frac{1}{3}$ (not shown) curve would correspond to a vertical line which passes through the instantaneous reheat point~\cite{CookEtAlReheating2015}. It is also to be noticed that in all the potentials treated within this formalism, the $\Tre$ converges around $10^{16} $GeV~\cite{DaiEtAlReheating2014,CookEtAlReheating2015,UenoEtAlConstraints2016}.

\section{Summary and Conclusion}\label{sec:conclusion}
In this \textit{Letter}, we have analyzed the reheating constraint on the K\"ahler moduli inflation within the slow-roll regime, following the formalism as in~\citep{CookEtAlReheating2015,UenoEtAlConstraints2016,DaiEtAlReheating2014}. Although there is no new observational constraint introduced in this reheating analysis as compared to ones trying to constrain standard inflation using $(n_s, r_T)$ plots~\cite{kabirreanalyzing2014}, embedding the eos parameter has enriched this analysis. That is the reason why this analysis can further constrain the allowed inflationary models. However as already stated, the details of single parameter potential, in general, are irrelevant if the analysis is done strictly within the slow-roll approximation. 

\section*{Acknowledgments}

RK thanks Jessica Cook for useful communication. 

\section*{References}

\bibliographystyle{elsarticle-num}
\bibliography{reheat,elspoly}

 \end{document}